# Reducing the efficiency droop by lateral carrier confinement in InGaN/GaN quantum-well nanorods

Chentian Shi,[1] Chunfeng Zhang,[1,a] Fan Yang,[1] Min Joo Park,[2] Joon Seop Kwak,[2,b] Sukkoo Jung,[3] Yoon-Ho Choi,[3] Xiaoyong Wang,[1] and Min Xiao[1,4,c]

[1]National Laboratory of Solid State Microstructures and Department of Physics, Nanjing University, Nanjing 210093, China
[2]Department of Printed Electronics Engineering, Sunchon National University, Sunchon, Jeonnam 540-742, Korea
[3]Emerging Technology Laboratory, LG Electronics Advanced Research Institute, Seoul 137-724, Korea
[4] Department of Physics, University of Arkansas, Fayetteville, Arkansas 72701, USA

**Abstract:**

Efficiency droop is a major obstacle facing high-power application of InGaN/GaN quantum-well (QW) light-emitting diodes. In this letter, we report the suppression of efficiency droop induced by density-activated defect recombination in nanorod structure of a-plane InGaN/GaN QWs. In the high carrier density regime, the retained emission efficiency in a dry-etched nanorod sample is observed to be over two times higher than that in its parent QW sample. We further argue that the improvement is a combined effect of the amendment contributed by lateral carrier confinement and the deterioration made by surface trapping.

[a] cfzhang@nju.edu.cn
[b] jskwak@sunchon.ac.kr
[c] mxiao@uark.edu

InGaN/GaN quantum wells (QWs) are perfectly suitable for demonstrating light-emitting diodes (LEDs) in the short-wavelength region.[1,2] However, their high-power applications have been hindered by an enduring issue of efficiency droop—the decrease in quantum efficiency of light emission with increasing carrier density.[3-7] To solve this problem, it is essential to avoid the leaky processes that reduce the emission efficiency at high carrier density. Some important progresses have been made on suppressing the efficiency droop in the past few years.[6-16] These advances have basically been achieved by meliorating the issues of current leakage[8-15] and Auger recombination.[16,17] Recently, another process of the density-activated defect recombination (DADR) has been identified to be also responsible for the efficiency droop in InGaN/GaN QWs.[18-20] Nevertheless, the way to avoid such efficiency droop has not been really investigated yet. Here, we propose that the effect of lateral carrier confinement in QW nanostructures can be employed to reduce this undesired DADR-induced efficiency droop.

The process of DADR decreases the emission efficiency with excess defect recombination at high carrier density as schematically shown in Figure 1a.[18-20] Upon increasing carrier density, the enhanced carrier scattering drives carriers to overcome the energy barriers and to populate the defect states.[18-20] In principle, such process can be suppressed if carrier motion can be confined in lateral directions by proper material/structure designs. To test this idea, we present a systematic optical study on a sample of InGaN/GaN nanorods in comparison with its parent of a-plane QWs. We have observed reduced efficiency droop in the nanorod sample and confirmed it as a result of lateral carrier confinement. We also argue that the extent of droop amendment made by lateral carrier confinement is harmed by the negative effect of surface trapping in the QW nanorods.

The nonpolar a-plane QW samples were grown on r-plane sapphire substrates consisting of a GaN buffer layer, an n-GaN layer, a 15 nm thick InGaN single-QW layer, and a p-GaN capping layer. The InGaN/GaN nanorods were fabricated with a dry-etching procedure as described in an earlier publication.[21] Second harmonic generation at 400 nm of ultrafast pulses from Ti:sapphire femtosecond laser was employed as excitation source for photoluminescence (PL) measurement. The emission was collected at the direction normal to the substrate and analyzed by a spectrograph (Sp 2500i, Princeton Instruments) equipped with a charge-coupled device cooled by liquid nitrogen. The excitation residual was eliminated by an ultrasteep long-pass filter (BLP01-405R-25, Semrock). The time-resolved PL (TRPL) spectrum was measured with the technique of time-correlated single-photon counting at a temporal resolution of ~ 50 ps provided by a fast single-photon avalanche diode (PDM, Picoquant) as described previously.[22] The PL lifetime was then extracted by fitting the decay component in the temporal window of first 10 ns with an exponential or biexponential decay function.

The scenario of lateral carrier confinement in a nanorod sample is depicted in Figure 1a. Localized states with potential minima decrease the possibility of defect recombination in InGaN QWs.[23-25] This effect of carrier localization induced by indium fluctuation ensures high quantum efficiency of bandedge emission.[24,25] The DADR is a type of carrier delocalization process which becomes dominant with increasing carrier density due to enhanced carrier scattering.[18,19] The carrier scattering drives the escape of carriers from localized states (Figure 1a) which recombine through defect states or other excess nonradiative centers, leading to an efficiency decrease of bandedge emission. Supposing the defect states are evenly distributed in space, the boundaries (Figure 1a, dashed green lines) of nanorods can physically block the channels linked between localized states inside the nanorods and defect states outside,[21,26] and, therefore, potentially amend the DADR-induced efficiency droop.

There are rapidly growing interests on optimizing InGaN LEDs with nanoarchitectured designs in the past few years, benefiting from some unique merits of nanostructures including strain relaxation and enhanced light extraction.[17,27-30] In this work, we carefully design the nanorod size to make sure that the procedure of nanofabrication mainly affects the process of DADR. We employ a parent sample of single InGaN/GaN QW grown on a nonpolar substrate, in which the DADR-induced efficiency droop has been identified very recently.[20] The average radius of QW nanorods (~ 130 nm, Figure 1b) is set to be in the same length scale as the carrier diffusion length in InGaN samples (60-500 nm).[31-33] This size is much larger than the Bohr radius (~ 3 nm) of excitons in InGaN samples,[34] so that the size effect on Auger recombination can be neglected. Here, we focus our study on the process of DADR by monitoring the correlation between efficiency droop and defect recombination.

We evaluate the efficiency droop by monitoring the integrated intensity of bandedge emission per unit excitation ($I_{Em}/I_{Ex}$) as a function of excitation fluence. The value of $I_{Em}/I_{Ex}$ is a metric that proportionally reflects the internal quantum efficiency of light emission from InGaN QWs as previously established in literature.[35,36] As a signature of efficiency droop, the dependence of $I_{Em}/I_{Ex}$ on the fluence develops from a "plateau" regime to a "decreasing" regime upon increasing excitation power (Figure 2a).[36] The efficiency droop is tightly associated with defect recombination where the intensity ratio between defect emission and bandedge emission ($I_D/I_B$) increases abruptly (Figures 2b & 2c). The peak of bandedge emission slightly shifts to the blue side due to state filling. These results confirm the presence of DADR-induced efficiency droop in the samples as discussed in an earlier work.[20] The saturation effect of defect states can be safely excluded here as the defect emission becomes much stronger with shorter-wavelength excitation.[20] We compare the experimental data recorded from the nanorod sample and its parent sample. Upon raising excitation fluence (> 0.2 µJ/cm$^2$), the retained efficiency is much higher in the nanorod sample, which means that the efficiency droop is partially reduced after nanofabrication (Figure 2a). The smaller value of $I_D/I_B$ in the nanorod sample indicates that such efficiency retention is realized with suppression of defect recombination.

To further identify the role played by lateral carrier confinement, we comparatively study the steady-state and transient PL emissions in the two samples. PL spectra recorded from the two samples are shown in Figure 3 under a sample fluence excitation (~ 0.1 µJ/cm$^2$). The emission spectra from both samples exhibit two bands with a blue bandedge emission and a yellow defect emission, respectively. In the nanorod sample, the light extraction is significantly enhanced with promoted bandedge emission (Figure 3). In spite of this, defect emission from the nanorod sample becomes weaker than that from the parent sample. This result can be well explained by the suppression of defect recombination with lateral carrier confinement in the nanorod sample, which is also evidenced by a peak blueshift of bandedge emission (Figure 3). The lateral carrier confinement induced by nanorod boundaries restrains carrier diffusion between localized states.[37,38] In this case, the possibility of carrier recombination through strongly localized states (low potential minimum) decreases, leading to the bandedge emission with higher photon energy.

The emission dynamics can provide more direct information about the carrier diffusion. In a QW sample, the carrier diffusion causes accumulation of carriers in strongly localized states, exhibiting a delayed-rise component following an abrupt rise in the TRPL spectrum.[21,26,37,38] Figure 4 shows the normalized TRPL spectra recorded from the nanorod sample and its parent sample under weak excitation (~ 0.1 μJ/cm$^2$). The amplitude of the delayed-rise component (D) is highlighted in comparison with the amplitude of overall signal (A). The amplitude ratio (D/A) in the nanorod sample is less than half of that in the parent sample, implying a strong confinement of the lateral carrier diffusion (Figure 1a). More importantly, this effect also blocks the channels of carrier escape from localized states to defect states, suppressing the process of DADR.

The above discussion has affirmed the potential to suppress the DADR-induced efficiency droop by incorporating nanorod structures in LEDs. However, a long-standing issue exists in such technology, i.e. surface states can be hardly avoided during nanofabrication.[29,39,40] The lifetime of PL decay in the nanorod sample is ~ 0.56 ns, which is very close to the value of ~ 0.58 ns in the parent sample (Figure 4). Such tiny difference suggests that the effect of surface states on emission dynamics is less important here than the cases of nanorods fabricated from c-plane QWs.[29,39,40] This result can be explained by the unique emission dynamics in a-plane samples benefiting from the absence of piezoelectric polarization.[41,42] The polarization field in c-plane samples causes carrier separation (quantum-confined Stark effect) that reduces the recombination rate.[34,43] Without polarization field, the electrons and holes in a-plane samples are better overlapped than that in c-plane samples.[41,42] In consequence, the carrier recombination are much faster in a-plane samples,[44,45] so that the trapping effect of surface states is less distinct in the TRPL spectrum recorded from a-plan QW nanorods. In spite of the negligible effect on emission dynamics, the surface states provide excess channels for carrier recombination (Figure 1a) which may limit the efficiency retention contributed by lateral carrier confinement. In other words, the procedure of dry etching has two contradictory effects on efficiency droop: suppression contributed by lateral carrier confinement and deterioration made by surface trapping. When the former factor becomes dominant, the efficiency droop is partially amended as we studied above.

To especially see the negative effect of surface trapping, we perform controlled experiments on c-plane QW samples. The control sample of multiple QWs consists of a GaN buffer, an undoped GaN layer, an n-type GaN layer, five pairs of 2.5 nm-thick InGaN QWs sandwiched between 13 nm GaN barriers and a p-type capping layer. The PL decay lifetime is generally much slower in the c-plane samples than in a-plane samples,[44,45] so that the surface effect can be more explicitly seen in the TRPL traces. The control samples are fabricated from a parent multiple QWs grown on a c-plane substrate. In comparison with a-plane samples studied above, the density of defect states is much lower with a weaker defect emission in the c-plane QW sample (Figure 5a, inset). The emission in the control QW sample decays much slower (~ 5.28 ns) (Figure 5a). In the nanorods (c-NR) fabricated from the c-plane QWs, the TRPL spectrum consists of two decay components. The lifetime parameters (amplitude ratios) of these two components are ~ 0.7 ns (~ 17 %) and ~ 4.85 ns (~ 83 %), respectively. The appearance of the faster component is an evidence of pronounced surface trapping effect (Figure 5a). The recombination of surface states is likely to be non-radiative since no significant defect emission is observed from the nanorod sample (Figure 5a, inset). The dependence of $I_{Em}/I_{Ex}$ on fluence indicates that, rather than being suppressed, the efficiency droop is deteriorated in the c-plane nanorod sample (Figure 5b). To check whether the result is a combined effect of lateral carrier confinement and surface trapping, we have further investigated another nanorod sample with surface passivation (c-NR-S). The surface passivation is realized by depositing a layer of $Al_2O_3$ on surface of the nanorod with the technique of atomic layer deposition. In the TRPL spectrum, the surface trapping component is not distinctly observable any more in the surface-passivated nanorod sample. The trace can be reproduced by an exponential decay function with a lifetime parameter of ~ 4.90 ns, suggesting that the density of surface states diminishes in this surface-passivated nanorod sample. The efficiency droop in the surface-passivated sample is significantly reduced as compared to that in the as-etched nanorod sample (Figure 5b), which confirms the negative role played by the surface states. From the above presented experimental evidences and discussions, we can safely conclude that the degree of suppressing the efficiency droop achieved by the lateral carrier confinement is harmed by the surface trapping.

In summary, we have found that the DADR-induced efficiency droop can be partially amended by lateral carrier confinement in QW nanorod structures. The full potential of this method may be approached by further reducing nanorod radius. However, the effect of surface trapping influences the efficiency retention contributed by lateral carrier confinement in nanorods. This drawback can be potentially solved by surface engineering with certain post-fabrication technology or more practically by employing epitaxial nano-LEDs with minimal density of surface states.[14,17,27,28,46] The technology studied here, together with the methods in literatures proposed on inhibiting other efficiency droop mechanisms,[6-16] can be intergraded for development towards droop-free LEDs, which is particularly desirable for high-power applications owing to the enhancement of light extraction in nanorods.

This work is supported by the Program of International S&T Cooperation (2011DFA01400, MOST), the National Basic Research Program of China (2013CB932903 and 2012CB921801, MOST), the National Science Foundation of

China (91233103, 61108001, 11227406 and 11021403), and NRF of Korea (K2011-0017325). The author C.Z. acknowledges financial support from New Century Excellent Talents program (NCET-09-0467), Fundamental Research Funds for the Central Universities, and the Priority Academic Program Development of Jiangsu Higher Education Institutions (PAPD). J.S.K acknowledges financial support from BK21 PLUS program in SCNU.

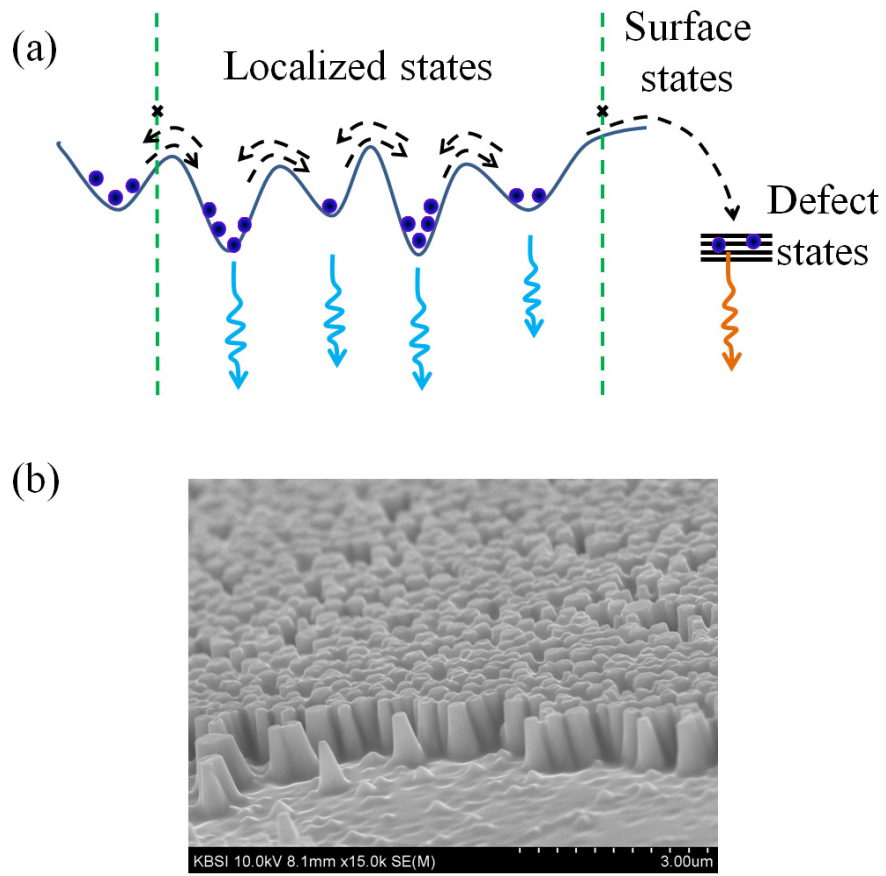

Figure 1. Lateral carrier confinement in QW nanorods. (a) Schematic sketch of the impact of lateral carrier confinement on the process of DADR (not in scale). The green dashed lines represents the boundaries of nanorods. (b) A SEM image of the nanorod sample. The average radius is ~ 130 nm.

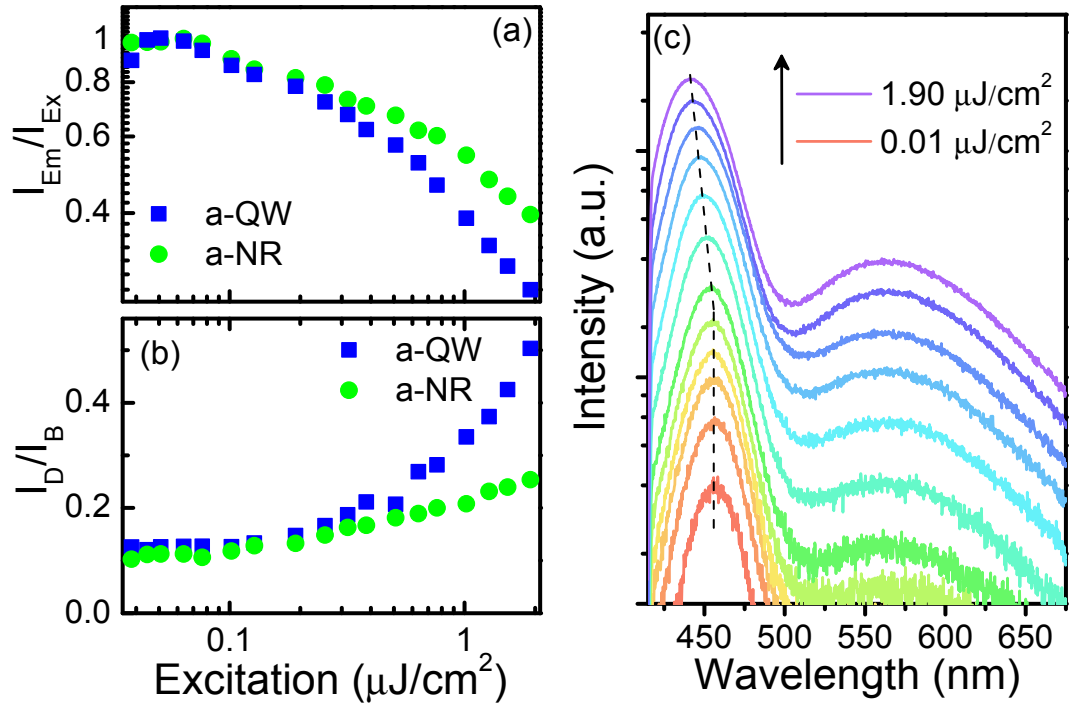

Figure 2. Reduced efficiency droop in the nanorod sample. (a) The normalized bandedge emission intensity per unit excitation power (logarithm scale) and (b) intensity ratio between defect emission and bandedge emission are plotted versus excitation fluence. The data from the nanorod sample and the parent QW sample are compared. (c) PL emission spectra from the nanorod sample recorded under different fluence excitation. The dashed line indicates the fluence-dependent shift of emission peak.

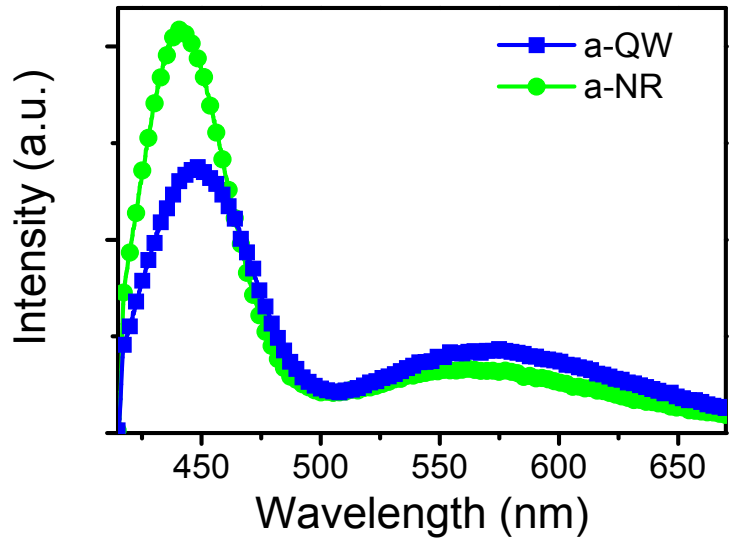

Figure 3. Time-integrated PL spectra from the nanorod sample and the parent sample recorded under same excitation (~ 0.1 μJ/cm$^2$).

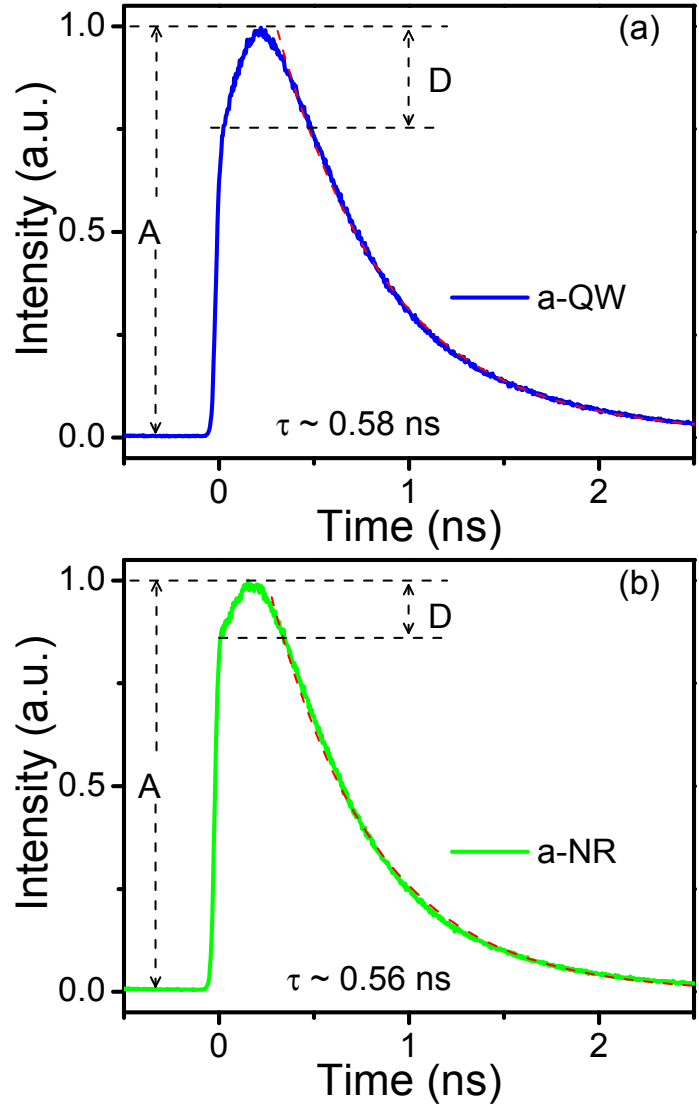

Figure 4. Transient optical evidence of lateral carrier confinement in nanorods. Normalized TRPL spectra recorded at the center wavelength of bandedge emission from (a) the parent sample and (b) the nanorod sample. Following the initial abrupt rise, the curves recorded from both samples exhibit delayed-rise components. The amplitude ratio between delayed-rise component (D) and overall signal (A) in the nanorod sample becomes less significant than that in the parent QW sample. The amplitude of delayed component was taken from the kink point of the rising edge in the TRPL curves. The read dashed lines are the curves fitting to an exponential decay function.

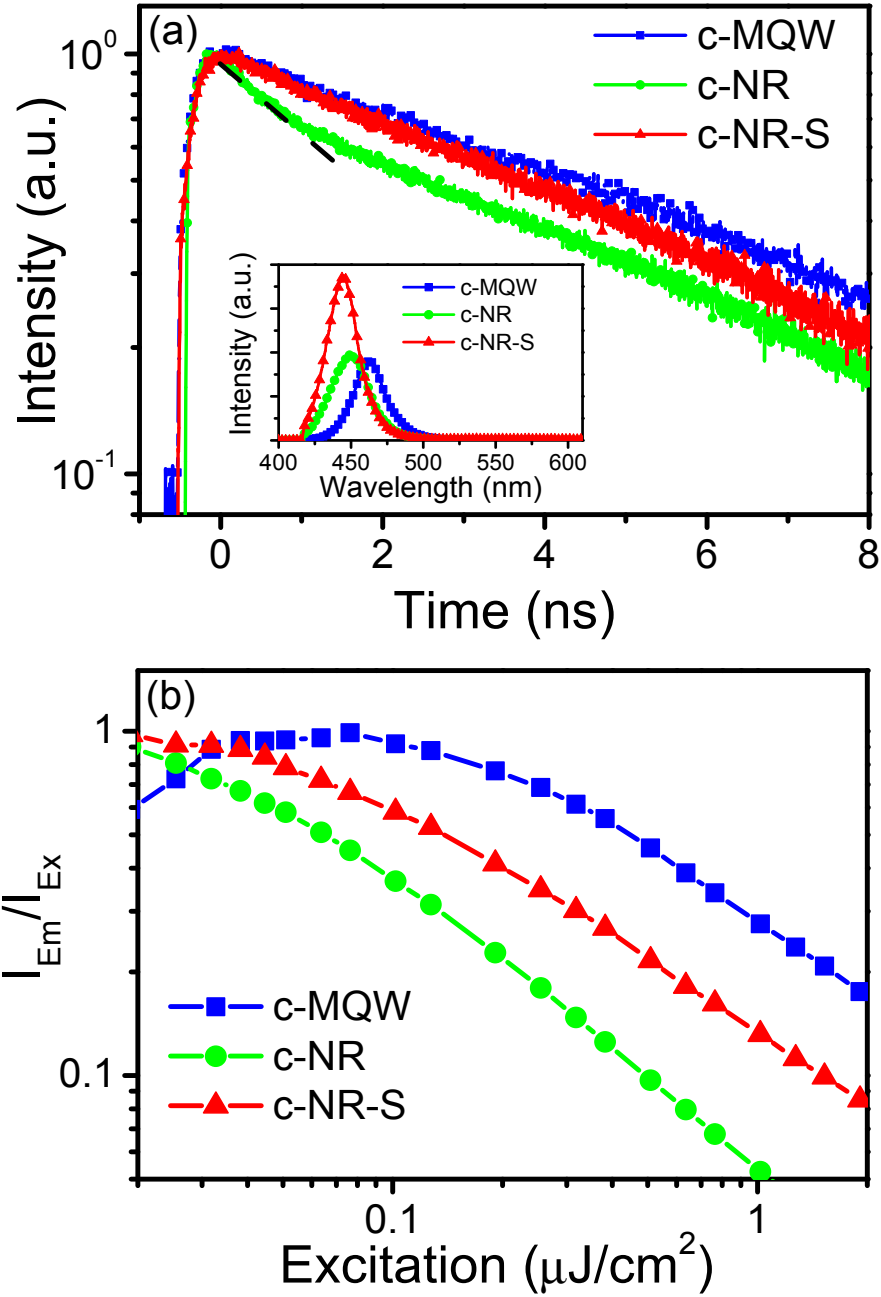

Figure 5. Control experiments on the surface trapping effect. Time-resolved (a) and time-integrated (a, inset) PL spectra of three control samples (the c-plane QWs (c-MQW), as-etched QW nanorods (c-NR), and surface-passivated QW nanorods (c-NR-S) recorded at the sample conduction are compared. The dashed line highlights the ultrafast decay component. (b) The normalized bandedge emission intensities per unit excitation power from the three samples are compared as functions of excitation fluence.